\documentclass[conference]{IEEEtran}
\usepackage{graphicx}
\usepackage{hyperref}
\usepackage{cite}
\usepackage{multirow}
\usepackage{array}
\usepackage{amsmath,amssymb,amsfonts}
\usepackage{algorithm}
\usepackage{algpseudocode}
\usepackage{amsmath}
\usepackage{graphicx}
\usepackage{textcomp}
\usepackage{adjustbox}
\usepackage{tabularx}
\usepackage{xcolor}
\usepackage{float}
\usepackage{enumerate}
\usepackage{booktabs}
\hypersetup{
    colorlinks=true,
    linkcolor=black,
    filecolor=magenta,      
    citecolor=black,
    urlcolor=black,
    pdfpagemode=FullScreen,
}
\def\BibTeX{{\rm B\kern-.05em{\sc i\kern-.025em b}\kern-.08em
    T\kern-.1667em\lower.7ex\hbox{E}\kern-.125emX}}
\begin{document}
\title{From Cracks to Crooks: YouTube as a Vector for Malware Distribution}
\author{\IEEEauthorblockN{Iman Vakilinia}
\IEEEauthorblockA{School of Computing, \\
University of North Florida,\\
Jacksonville, FL, USA\\
i.vakilinia@unf.edu}
}

\maketitle

\begin{abstract}

With billions of users and an immense volume of daily uploads, YouTube has become an attractive target for cybercriminals aiming to leverage its vast audience. The platform's openness and trustworthiness provide an ideal environment for deceptive campaigns that can operate under the radar of conventional security tools.
This paper explores how cybercriminals exploit YouTube to disseminate malware, focusing on campaigns that promote free software or game cheats. It discusses deceptive video demonstrations and the techniques behind malware delivery. Additionally, the paper presents a new evasion technique that abuses YouTube’s multilingual metadata capabilities to circumvent automated detection systems. Findings indicate that this method is increasingly being used in recent malicious videos to avoid detection and removal.
    
\end{abstract}

\section{Introduction}

In recent years, cybercriminals have increasingly leveraged mainstream social media platforms as vectors for malware distribution. Among these, YouTube has emerged as a particularly attractive target due to its massive global user base and high levels of user engagement. With over 2.7 billion active users in 2024 \cite{YT}, YouTube serves as a trusted platform for entertainment, education, and software tutorials making it a fertile ground for exploitation. Attackers have capitalized on this trust to deliver malware through deceptive videos, often bypassing traditional detection mechanisms employed by hosting providers and endpoint security solutions.
A prevalent tactic involves hijacking legitimate YouTube channels, often with existing subscriber bases, to lend credibility to malicious content. Once compromised, these channels are used to upload videos that advertise free cracked versions of commercial software, video games, or cheat tools. These videos typically promise access to premium or hard to find applications at no cost, especially to younger or less technically aware audiences. The malware payload is delivered through links embedded in the video description or comment section, often disguised as legitimate file sharing URLs \cite{yamagishi2024users}.

Furthermore, the emergence of Malware-as-a-Service (MaaS) has significantly lowered the barrier to entry for cybercriminals, enabling even unskilled actors to participate in sophisticated malware distribution campaigns. MaaS refers to a business model in which malware developers offer ready-made malicious tools and infrastructure for rent or sale on underground forums and darknet markets. These services often include not only the malware binaries themselves, but also customizable builders, control panels, and technical support \cite{DN}. This model plays a crucial enabling role in the abuse of platforms like YouTube for malware dissemination. 
In many observed campaigns, attackers distribute information stealer malware, a class of malware designed to extract browser credentials, cryptocurrency wallets, session cookies, and other sensitive data from infected systems. Recent studies have documented the proliferation of stealers such as RedLine, Lumma, Raccoon, and Vidar via these deceptive campaigns \cite{yucrackedcantil,IS,FN, bang2025threat}.
These stealers often come with a builder interface that allows the attacker to quickly generate a customized payload, specify a command-and-control (C2) address, and select the types of data to exfiltrate.

With access to these tools, cybercriminals can rapidly deploy new campaigns. They simply need to upload a convincing video on a hijacked or newly created YouTube channel, embed a malicious link in the description or comments, and provide basic social engineering often promoting free software cracks or game cheats \cite{yamagishi2024users}. The malicious file is usually wrapped in a password-protected ZIP or RAR archive to evade detection by antivirus software and hosting providers.
Moreover, the MaaS economy fosters a rapid iteration cycle. Once a payload is flagged by antivirus engines, the attacker can regenerate a new variant using the same builder. This dynamic adaptability, combined with YouTube’s high-trust environment, creates a powerful vector for mass scale malware distribution. MaaS decouples technical sophistication from operational execution, allowing a wide range of threat actors from amateurs to organized cybercrime groups to abuse platforms like YouTube at scale. This highlights the importance of not only securing endpoint systems, but also monitoring platform level abuse and disrupting the underlying malware supply chain.


This paper investigates the techniques employed by cybercriminals to distribute malware via YouTube, particularly through the promotion of free software downloads or game cheats. It examines deceptive video demonstration strategies, as well as the underlying architecture used to deliver malware to victims. Furthermore, the paper introduces a novel evasion method that leverages YouTube's multilingual metadata feature to bypass automated detection systems. Analysis reveals that a significant number of recent malicious videos on the platform are utilizing this technique to evade identification and takedown mechanisms.

\section{Cybercriminals Techniques}

A key factor in the success of malware distribution through YouTube lies in the attackers’ ability to convincingly simulate legitimacy. Unlike traditional phishing campaigns that rely heavily on urgency or fear, these YouTube-based malware campaigns are often engineered to build trust, exploiting the high perceived credibility of video content and social proof indicators.

\subsection{Deceptive Video Demonstrations}
\begin{figure*}[t]
    \centering
    \includegraphics[width=0.9\textwidth]{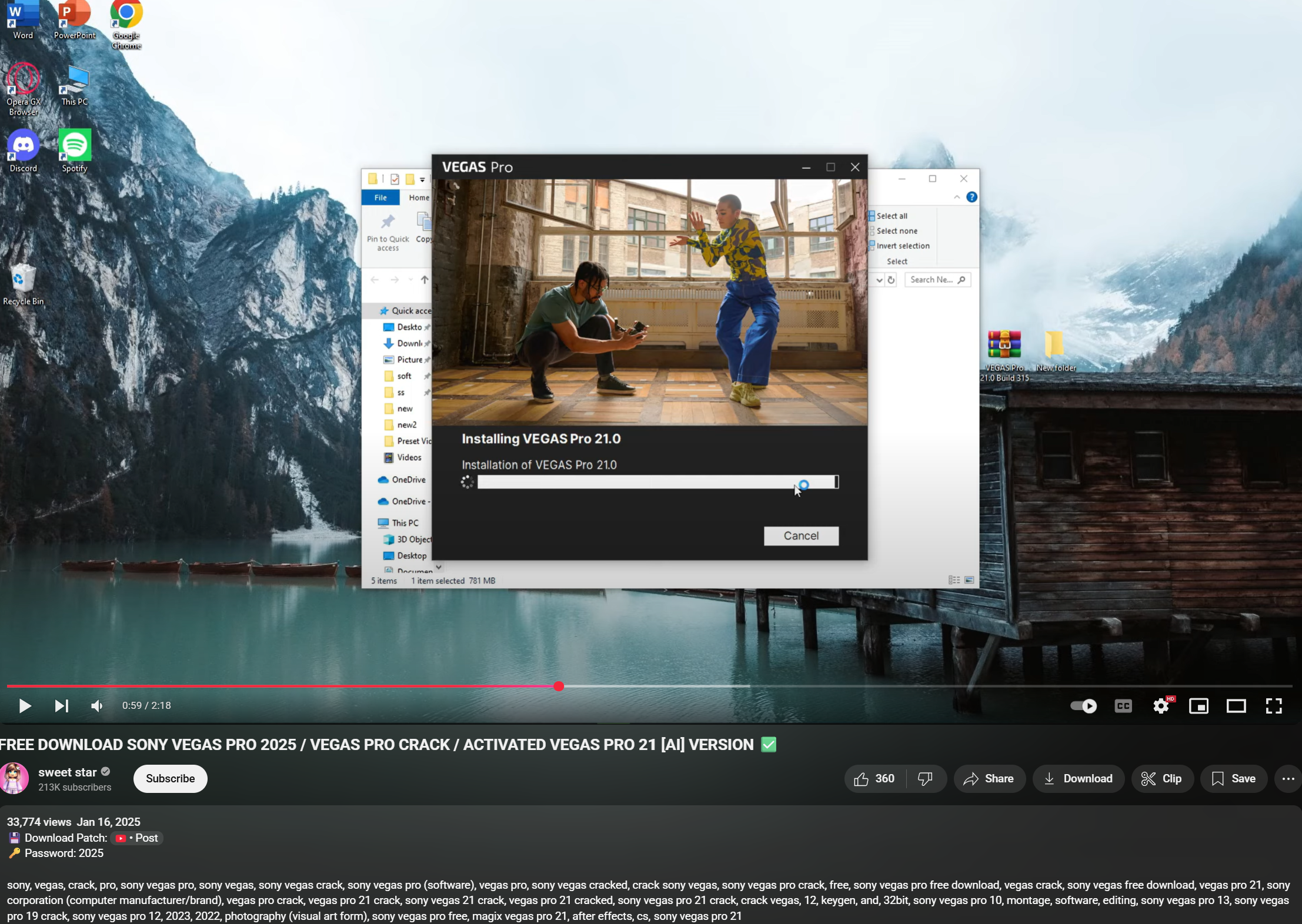}
    \vspace{-0.1in}
    \caption{A fake YouTube video advertising free software for malware distribution}
    \vspace{-0.1in}
    \label{youtube}
\end{figure*}

Attackers commonly upload videos that simulate the installation and successful use of the advertised software, which is often a cracked version of popular applications, games, or cheating tools. These videos typically include screen recordings showing the file being opened and installed without issue, and then the application launching and appearing to function as expected. Figure \ref{youtube} shows a sample of such a video.

This visual demonstration plays a critical psychological role. It gives viewers a false sense of validation, reinforcing the impression that the file is safe and functional. In many cases, attackers go to great lengths to script and edit the video to avoid displaying anything suspicious, such as system warnings, antivirus flags, or broken software behavior.

To further build credibility, attackers seed the comment section with fake user testimonials, such as: 

\begin{itemize}
    \item It worked! Thanks!
    \item Finally got it running!
    \item No virus, just follow the steps!
    \item Can’t believe this actually works!
    \item Been looking for this everywhere!
    \item 100\% legit
    \item Works like a charm!
    \item Thanks bro, saved my day
    \item I was skeptical but it’s real
    \item Just downloaded and it's perfect!
\end{itemize}

These comments are typically short, enthusiastic, and designed to simulate authentic user experiences. The accounts posting such comments are likely either compromised legitimate user accounts, newly created, or in some cases, operated by bots. This tactic is commonly used in social engineering campaigns to build trust among unsuspecting viewers and increase the likelihood of engagement with malicious content.

Many of the malicious videos show high numbers of views, likes, and seemingly organic engagement. Attackers may use tools to inflate the view and likes count. 

These inflated metrics serve as heuristics of legitimacy. Most users subconsciously associate high view counts and likes with safe, trustworthy content. When combined with the fake comments and a well-crafted video demonstration, the deception becomes highly convincing.

\subsection{Leveraging Hijacked YouTube Channels}

Attackers hijack legitimate YouTube accounts, then repurpose them to host malware-distributing content. These accounts may have thousands of preexisting subscribers, further boosting the perceived authenticity of the video and the channel itself. The use of older, verified, or monetized accounts also helps bypass some of YouTube’s newer restrictions on video uploads or links.

\subsection{Link Obfuscation and Hierarchical Redirection via YouTube Posts}

An increasingly common technique among malware distributors on YouTube involves the use of intermediary redirection layers, often implemented through YouTube ``Community Posts" hosted on other compromised or hijacked channels. This approach not only enhances the perceived legitimacy of the campaign, but also serves as a dynamic control point for managing malicious links more flexibly and evasively.

In this technique, instead of placing the direct malware link in the description or pinned comment, the video contains a link to a YouTube community post hosted on another hijacked account.
The target YouTube post was often originally published as benign content which the attacker subsequently edits to include the malicious download link. This technique provides multiple advantages:

\begin{itemize}
    \item First, it evades YouTube’s automated detection, which may not rescan edited content.
    \item Second, it preserves social trust signals, such as existing likes or user engagement from before the post was hijacked or altered.
    \item Third, it exploits platform limitations, as YouTube does not currently provide users with a way to report or flag posts, unlike videos.
    \item Fourth, by directing users to a common intermediary post, attackers can manage multiple malware campaigns more efficiently. Several videos spread across different hijacked channels can link to a single community post, creating a centralized redirection hub. This redirection layer allows attackers to regularly update the final download URLs or file-hosting services without re-uploading or altering the video description/comment itself. This is critical because malware hosting domains, IP addresses, or file hashes are often quickly flagged and blacklisted by antivirus vendors or hosting platforms. This hierarchical setup increases the resilience of the campaign. Even if some videos are taken down, the central post remains active, and vice versa. The attacker can also reassign videos to new posts or rotate between different intermediary layers.
\end{itemize}

\subsection{Malware Hosting Platforms}
In the majority of observed cases, attackers distribute their malware payloads via mainstream file sharing platforms, most notably MediaFire, which is frequently used to host the encrypted archive files containing malware. In a smaller subset of cases, services such as Dropbox have also been leveraged for the same purpose. These platforms are commonly chosen due to their widespread trust, free hosting options, and lenient content monitoring, which together help reduce user suspicion and delay detection.

To evade blacklisting by antivirus vendors and takedown by hosting providers, attackers frequently update or rotate the hosted malware archive typically within a 48-hour window. These updates may include changing the download URL, using newly compiled malware variants, employing different password-protected ZIP or RAR files, or relocating the payload to alternative hosting platforms or domains. In addition to modifying the delivery mechanisms, attackers may also alter the internal configuration of the malware itself, such as updating the IP address or domain of the command-and-control (C2) server, changing hardcoded credentials, or adjusting the dropper’s behavior to bypass detection and adapt to evolving defensive measures.
This constant rotation strategy significantly reduces the lifespan of static detection signatures and frustrates both researchers and automated defenses.

\begin{figure}[t]
    \centering
    \includegraphics[width=0.45\textwidth]{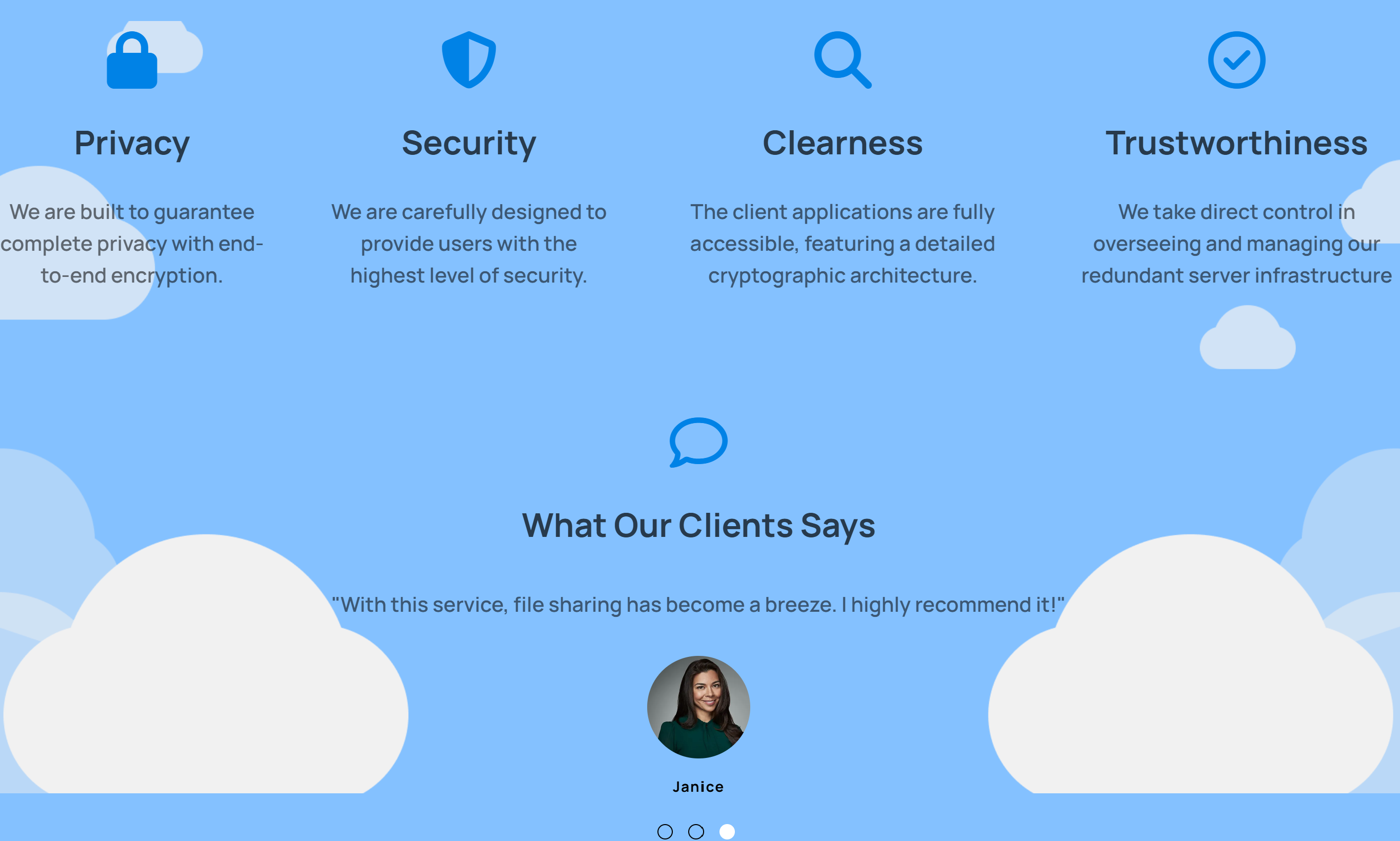}
    \vspace{-0.1in}
    \caption{Fake file hosting platform website for malware distribution}
    \vspace{-0.1in}
    \label{file hosting}
\end{figure}

Beyond using public file sharing services, attackers have also been observed setting up custom websites to host malware.
Some attacker controlled websites are crafted to mimic the appearance and functionality of legitimate file hosting services as shown in figure \ref{file hosting}. These sites often include deceptive elements such as download buttons, fake virus scan verifications, upload dates, fake user ratings, and password hints, all designed to increase credibility and lure victims into downloading the malicious payload. These deceptive visual cues increase trust and encourage users to download the payload without hesitation.

\begin{figure}[t]
    \centering
    \includegraphics[width=0.45\textwidth]{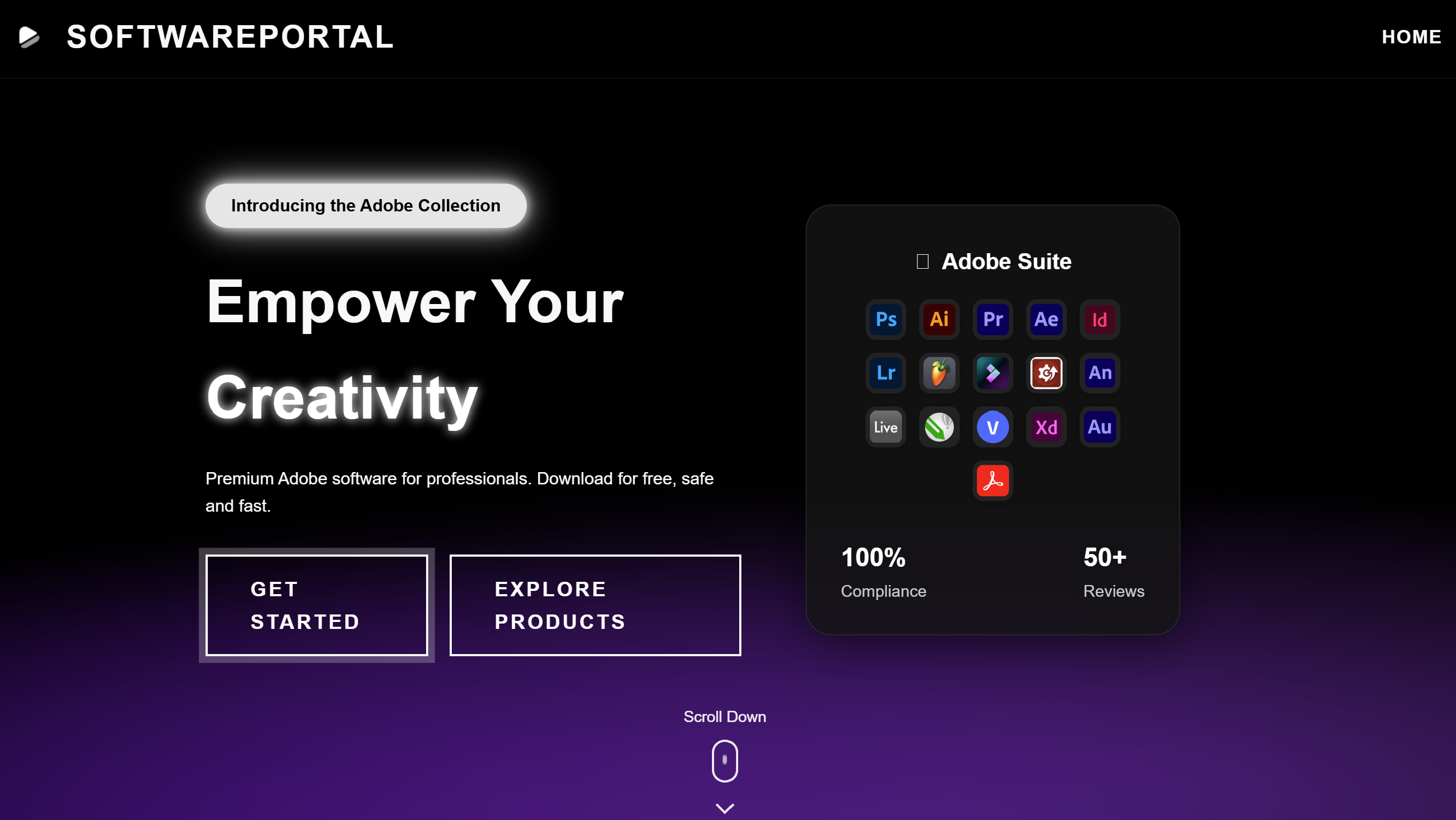}
    \vspace{-0.1in}
    \caption{Fake software portal website for malware distribution}
    \vspace{-0.1in}
    \label{software portal}
\end{figure}

\begin{figure}[t]
    \centering
    \includegraphics[width=0.45\textwidth]{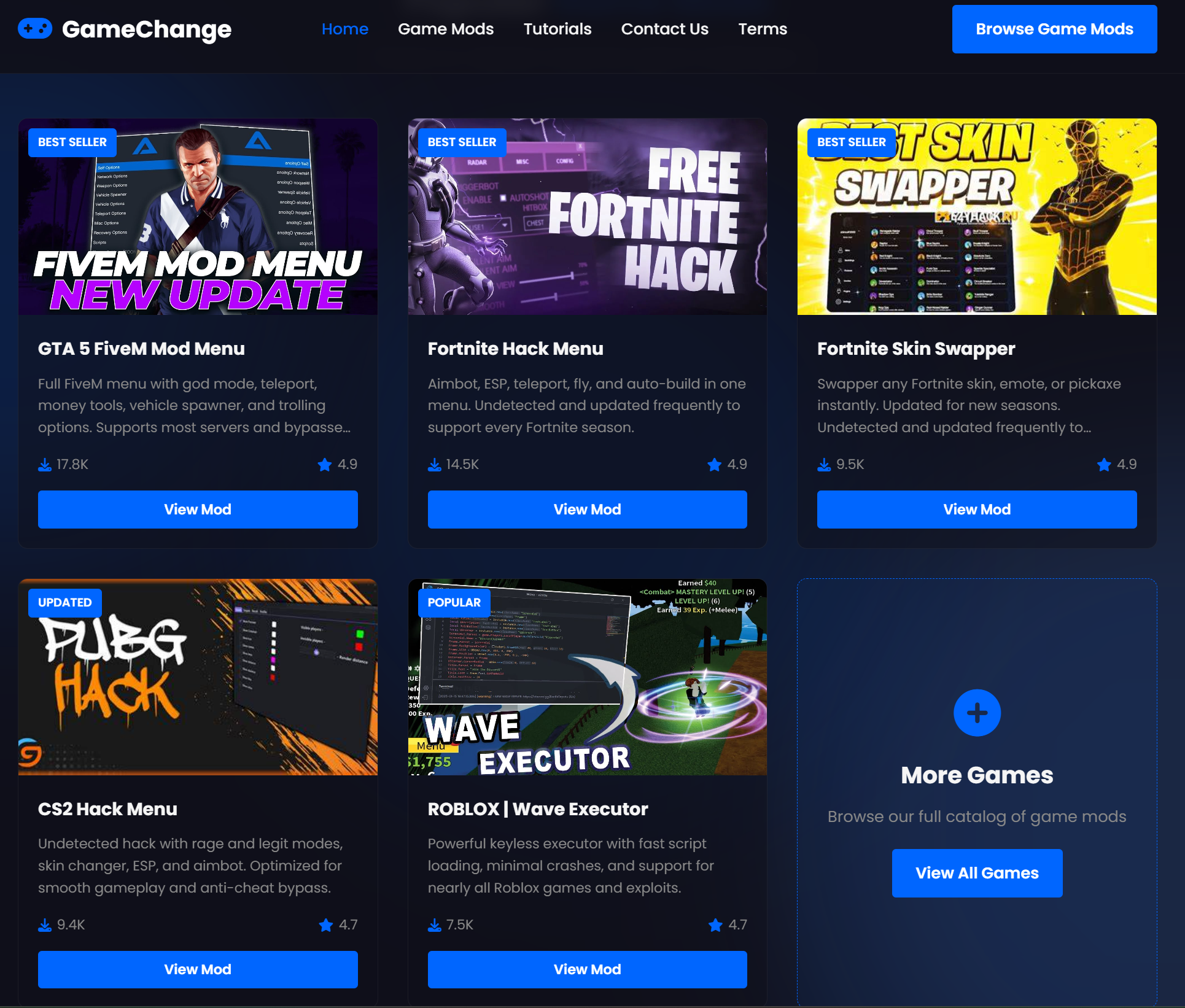}
    \vspace{-0.1in}
    \caption{Fake game portal website for malware distribution}
    \vspace{-0.1in}
    \label{game portal}
\end{figure}

In other cases, attackers build websites that resemble genuine software distribution or gaming portals as shown in figures \ref{software portal} and \ref{game portal}. These sites may offer fake cracked versions of popular software or game cheats, use recognizable branding, and present fake user reviews or star ratings. This adds another layer of social engineering and legitimizes the download process.

A notable technical choice by attackers is the consistent use of Cloudflare as a reverse proxy and content delivery network (CDN) for their malicious websites. Cloudflare offers a free tier with many advanced features and a simple setup process, making it accessible to even low-skilled threat actors.
Cloudflare provides the following services that benefit attackers:

\begin{itemize}
    \item IP Masking: Cloudflare hides the true origin IP address of the server hosting the malicious content, making it difficult for researchers or authorities to locate and take down the backend infrastructure.

    \item SSL Certificates: Cloudflare provides free HTTPS encryption, which can make the site appear more trustworthy to victims who look for a padlock icon before downloading.

    \item High Availability: Cloudflare's global network ensures that malicious content can be delivered quickly and reliably across different regions. 

\end{itemize}

In essence, Cloudflare is being abused as a shield, offering attackers the same protections designed to benefit legitimate web services.

Analysis of domain registrations for some of the attacker controlled sites reveals that they have WHOIS information redacted for privacy. Attackers deliberately choose domain registrars that offer WHOIS privacy protection, preventing public access to registrant data such as name, email address, and phone number. This tactic significantly hinders attribution efforts and makes it difficult to connect multiple malicious domains to a single actor or group.

\subsection{Malware characteristics}

To bypass antivirus engines and hosting provider scans, the malware payload is typically embedded within a password-protected ZIP or RAR file. The password is usually mentioned in the description or a pinned comment.
This not only helps evade automatic file scanning, but also enhances the illusion of exclusivity or protection making it seem like the software is protected for legitimate reasons. Once opened or decompressed, the malware archive displays a variety of folders and files including .dat, .txt, .xml, and .dll formats alongside the main executable. This structure is deliberately crafted to mimic the layout of legitimate software and reduce user suspicion.

The malware samples analyzed in this study exhibit multiple evasion techniques designed to hinder detection and analysis. Notably, the file size of the malware is intentionally inflated often exceeding the upload limits of certain antivirus scanning services such as VirusTotal to avoid static analysis. Additionally, the installer is heavily packed and obfuscated, employing layers of encryption and code manipulation to complicate reverse engineering efforts.

A common feature observed across samples is the ability to detect execution in sandboxed or virtualized environments. When such environments are detected commonly used by malware analysts and automated analysis tools, the malware refrains from executing its malicious payload, thereby avoiding behavioral detection.

A detailed VirusTotal analysis of one such sample is available at the following reference \cite{VT}, and a dynamic analysis using ANY.RUN can be found here \cite{AR}. As of the time of writing, 50 out of 72 antivirus engines on VirusTotal have flagged the sample as malicious.

All the observed malware samples appear to be variants of well-known information stealers, such as Lumma. A more comprehensive technical analysis of similar malware samples is provided by Yu in \cite{yucrackedcantil}.

\subsection{Multilingual Misrepresentation Technique}
The author has identified a novel technique employed by attackers to evade detection mechanisms on YouTube. YouTube provides an API (the YouTube Data API v3) which allows developers to programmatically interact with YouTube’s infrastructure. Security scripts often use this API to scan video metadata (e.g., titles and descriptions) for suspicious keywords such as “free,” “download,” “crack,” or names of specific software and games, in an effort to flag potentially malicious content \cite{yamagishi2024users}.

However, the author observed a discrepancy between the metadata returned by the API and what is actually displayed on the YouTube website. Specifically, while the API returned benign text devoid of any suspicious indicators, the same video when viewed in a browser displayed a title and description suggesting access to cracked or pirated software.

\begin{figure*}[t]
    \centering
    \includegraphics[width=0.9\textwidth]{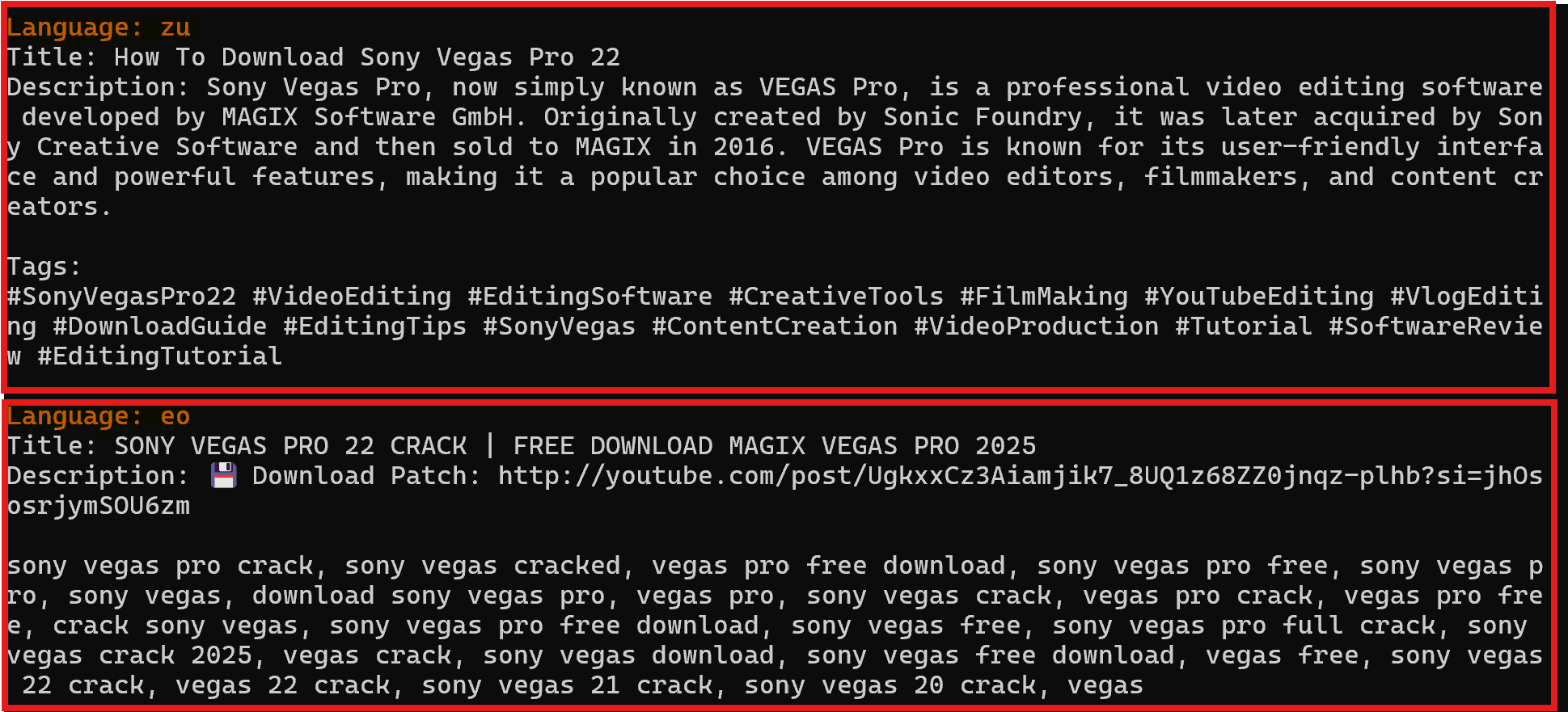}
    \vspace{-0.1in}
    \caption{Upper red box is the benign text in default language, and below red box is the malicious description.}
    \vspace{-0.1in}
    \label{multiling}
\end{figure*}

Upon further investigation, the author discovered that attackers were exploiting YouTube’s multilingual metadata support. YouTube allows creators to provide titles and descriptions in multiple languages. In this evasion technique, attackers set the default language to a less commonly used language (e.g., Zulu) to contain harmless text, while using English or other widely spoken languages to present the actual malicious content. Since YouTube's web interface automatically displays metadata in the viewer’s preferred language (e.g., English), the malicious content is revealed during normal browsing. In contrast, the YouTube Data API when queried without specifying a language returns only the default benign metadata. This discrepancy allows malicious videos to evade detection by automated systems that rely solely on API data and do not account for language-specific localizations. Figure \ref{multiling} presents an example of a malicious YouTube video where attackers exploit multilingual metadata fields to embed benign content (e.g., in the default language) alongside malicious description.

The author has responsibly disclosed this behavior to the Google security team.

\section{Malicious Videos With Multilingual Misrepresentation}

To assess the prevalence of this attack technique, a script was developed using the YouTube Data API to query publicly available videos. The search was constrained to videos published in 2025 that included keywords such as ``free," ``download," ``crack," and the name of specific software or games. To mitigate the effects of the Multilingual Misrepresentation technique, the search was explicitly configured to target English language metadata. To minimize the strain on YouTube's infrastructure, an exhaustive and resource-intensive search was deliberately avoided. However, it is reasonable to assume that a more comprehensive search would uncover a higher number of malicious videos.

Table \ref{stats} summarizes the number of videos identified as potentially involved in malware distribution, along with the total view counts for videos related to each software or game, as of July 19, 2025. These numbers reflect the popularity of certain applications among users seeking unauthorized or pirated versions.
In total, 175 videos were identified, accumulating a combined view count of 3,531,008.
As part of responsible disclosure, the author has shared the list of identified malicious videos employing the Multilingual Misrepresentation technique with the Google security team.

\begin{table}[h!]
\centering
\begin{tabular}{|c|c|c|}
\hline
 \textbf{Software/Game} & \textbf{Number of Videos} & \textbf{Total Views} \\
\hline
Adobe Photoshop & 17 & 202,076
 \\
\hline
Adobe Illustrator & 17 & 98,957
 \\
\hline
Adobe Premiere Pro & 8 & 363,328
 \\
\hline
Adobe After Effects & 9 & 22,368
 \\
\hline
Adobe InDesign & 12 & 16,204
 \\
\hline
Adobe Acrobat Pro & 7 & 63,274
 \\
\hline
Vegas Pro & 9 & 21,318
 \\
\hline
Camtasia Studio & 3 & 2,164
 \\
\hline
CorelDRAW Graphics Suite & 8 & 88,330
 \\
\hline
Filmora & 7 & 73,810
 \\
\hline
FL Studio & 16 & 603,559
 \\
\hline
Ableton Live & 5 & 13,022
 \\
\hline
DaVinci Resolve Studio & 2 & 7,334
 \\
\hline
AutoCAD & 6 & 21,932
 \\
\hline
Fortnite & 7 & 7,258
 \\
\hline
Valorant & 13 & 13,583 
\\ 
\hline
Roblox & 35 & 1,913,491
 \\
\hline
\end{tabular}
\vspace{+0.1in}
\caption{Statistics of Malicious YouTube Videos for Malware Distribution}
\label{stats}
\end{table}

Yamagishi et al. \cite{yamagishi2024users} conducted a detailed measurement study of malicious YouTube videos used for malware distribution during the period from December 20, 2023, to April 30, 2024. However, their research does not appear to address the multilingual misrepresentation technique, which seems to have emerged more recently as a method used by attackers to evade detection.

\section{Discussion}

One of the critical challenges in combating malware distribution through platforms such as YouTube is the lack of widespread user awareness and education on cybersecurity best practices. Many users, particularly younger audiences or those less technically informed, remain vulnerable to deceptive social engineering tactics, such as promises of free software, game cheats, or exclusive digital content. Raising public awareness and improving digital literacy is crucial in reducing the effectiveness of such campaigns.

From the platform’s side, YouTube provides a mechanism for reporting potentially harmful or malicious content. However, based on the author’s direct experience, this process is often inefficient and lacks the urgency warranted by the severity of the threat. For example, although the author received an acknowledgment email from YouTube after reporting multiple malicious videos, the flagged content remained publicly accessible for a long time. Considering the relatively straightforward nature of identifying videos that promote known malware delivery schemes, this delay is concerning. Malicious videos can accumulate thousands of views in a short period of time, significantly increasing the risk of infection before intervention occurs.

A related issue is YouTube’s historically weak response to other forms of cybercrime, most notably large-scale cryptocurrency scams. In multiple high-profile incidents, cybercriminals have hijacked legitimate YouTube channels to host live streamed videos impersonating major tech figures or companies (e.g. Elon Musk, Tesla, Ethereum Foundation), promoting fraudulent crypto giveaways \cite{vakilinia2022cryptocurrency}. Despite widespread media coverage and user complaints, YouTube has often been criticized for failing to respond quickly or effectively. This precedent raises broader concerns about the platform's capability and willingness to handle cyber threats proactively and at scale.

These challenges may arise from technical limitations in automated detection systems, underinvestment in content moderation resources, or prioritization of other forms of abuse over cybersecurity concerns. Furthermore, there appears to be a regulatory gap. Current laws and industry standards do not impose binding obligations on content platforms to act within a specific time frame or to report remediation results transparently. This lack of accountability enables attackers to exploit the platform repeatedly with limited consequences.

To address these issues, a comprehensive strategy is required. This should include:

\begin{itemize}
    \item Enhanced user education to promote safer online behavior,
    \item Stronger automated detection techniques that can identify multilingual and evasive abuse patterns,
    \item Faster and more transparent moderation workflows.
    \item Clearer regulatory frameworks that establish enforceable requirements for timely action and reporting by platforms in response to cyber threats.
\end{itemize}

Without coordinated action across technical, educational, and policy domains, platforms like YouTube will continue to be exploited by cybercriminals for malware delivery, fraud, and other harmful activities.

\section{Conclusion}

This paper investigated the use of YouTube as a vector for malware distribution. By analyzing attacker behavior, video content, and distribution tactics, we identified a novel evasion method termed multilingual misrepresentation, in which malicious actors embed benign metadata into less commonly used language fields while reserving malicious titles and descriptions for widely spoken languages such as English. This technique allows videos to bypass detection mechanisms that rely solely on default API responses, which often return only the default language entry.

To assess the prevalence of this technique, we developed a targeted script using the YouTube Data API and conducted an empirical analysis of videos published in 2025 that advertised free or cracked versions of popular software and games. Our investigation uncovered 175 videos that actively use multilingual metadata to distribute malware. These videos collectively pose a significant threat to unsuspecting users, especially given the added credibility conferred by hijacked channels and fake user comments.

In line with responsible disclosure practices, the details of the identified videos and the associated evasion technique have been reported to the Google Security Team. We hope that this research contributes to improved detection mechanisms on content sharing platforms and raises awareness about the evolving strategies used by malware distributors. 

\bibliographystyle{IEEEtran}
\bibliography{refs}
\end{document}